\documentclass[preprint]{sig-alternate}
\usepackage{float}
\usepackage{subfigure}
\usepackage{microtype}
\usepackage{tikz}
\usepackage{graphicx}
\usetikzlibrary{calc,positioning}

\begin{document}

\setcopyright{acmcopyright}

\doi{}

\isbn{}

\global\copyrightetc{\vspace{2mm}\footnotesize

To appear in \textit{Communications of the ACM}, June 2017, pp.~72-80.

\vspace{2mm}
\copyright \the\copyrtyr\ ACM \the\acmcopyr
\vspace{1mm}}



%

\title{Learnable Programming: Blocks and Beyond}

 \numberofauthors{5}
%
\author{
%
%
\alignauthor
David Bau\\
\affaddr{MIT CSAIL}\\
\affaddr{Cambridge, MA USA}\\
\email{davidbau@csail.mit.edu}
\alignauthor
Jeff Gray\\
\affaddr{Dept. of Computer Science}\\
\affaddr{University of Alabama}\\
\affaddr{Tuscaloosa, AL USA}\\
\email{gray@cs.ua.edu}
\alignauthor
Caitlin Kelleher\\
\affaddr{Dept. of Computer Science}\\
\affaddr{\sloppy Washington University}\\
\affaddr{St. Louis, MO USA}\\
\email{ckelleher@cse.wustl.edu}
\and  
\alignauthor
Josh Sheldon\\
\affaddr{MIT CSAIL}\\
\affaddr{Cambridge, MA USA}\\
\email{jsheldon@csail.mit.edu}
\alignauthor
Franklyn Turbak\\
\affaddr{Computer Science Dept.}\\
\affaddr{Wellesley College}\\
\affaddr{Wellesley, MA, USA}\\
\email{fturbak@cs.wellesley.edu}
}

\maketitle
\begin{abstract}
Blocks-based programming has become the lingua franca for introductory coding. Studies have found that experience with blocks-based programming can help beginners 
learn more traditional
text-based languages.
We explore how blocks environments improve learnability
for novices by 1) favoring recognition over recall, 2) reducing cognitive load, and 3) preventing errors. Increased usability of blocks programming has led to widespread adoption within introductory programming contexts across a range of ages. Ongoing work explores further reducing barriers to programming, supporting novice programmers in expanding their programming skills, and transitioning to textual programming. New blocks frameworks are making it easier to access a variety of APIs through blocks environments, opening the doors to a greater diversity of programming domains
and supporting greater experimentation for novices and professionals alike.

\end{abstract}

%
%


%
%

%
%



\section{Introduction}

\begin{figure}
\centering
\fbox{\includegraphics[width=0.97\columnwidth]{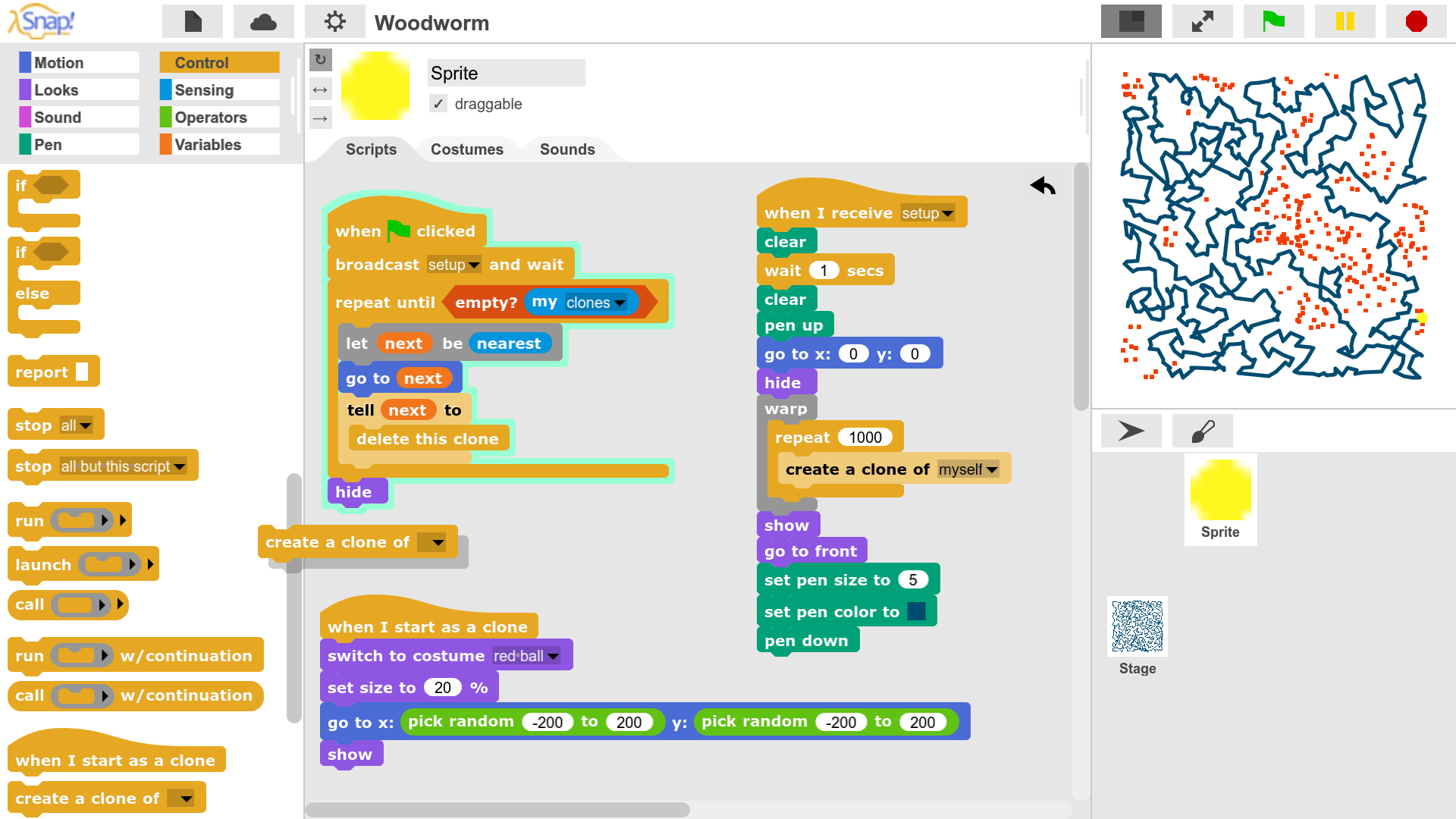}}
\caption{Snap!~is an example of a blocks-based programming environment. Users drag blocks from a palette of programming elements (left) into a workspace (center), where blocks can be assembled into programs.  Snap!~also provides an output window (top-right) and a sprite picker (bottom-right).}
\label{snapblocks}
\end{figure}

A global push to broaden participation in computer science has led to an explosion of interest in blocks-based programming. Visual blocks are used by more than a dozen programming tools (see the Sidebar below). 
Millions of students receive their first exposure to programming via these tools in
courses and activities like Code.org's
\textit{Hour of Code}.
Blocks allow beginners to compose programs without struggling with the frustrations of syntax (Figure \ref{snapblocks}).

There is increasing interest in developing and studying blocks languages. At VL/HCC 2015, a small workshop session called \textit{Blocks and Beyond}\footnote{\url{http://cs.wellesley.edu/blocks-and-beyond}} ballooned to a large event, with 51 submissions and 36 presenters.
Researchers shared work in new blocks languages, interface innovations, 
domain-specific applications of blocks, 
and ways to make blocks languages more effective and accessible for diverse coders.

This article explores how blocks impact the learnability of programming. We begin by reviewing studies on the effectiveness of blocks languages. Then we discuss the key features of blocks languages and how they relate to learning. Finally, we look at applications of blocks in new domains and discuss tools for creating  your own blocks language.

\section{Do Blocks Languages Work?}

Watching beginners create their first programs with blocks can be simultaneously inspiring and unsettling. Empowered by blocks, novices will rapidly build complex, often delightful creations. But just as quickly, they fill their screen with clumsy and intricate code \cite{meerbaum2011habits}.
A seasoned programmer inspecting a beginner's disordered assembly might worry that snapping together colorful blocks has nothing to do with ``real code.'' 
But what is ``real code,'' and why learn it?

\subsection{What is ``Real Code?''}

The purpose of a blocks-based tool is to make programming easy to learn. But programming education can have two distinct endpoints:

\begin{enumerate}
\item Development of expertise in professional programming.
\item Ability to accomplish other goals by creating programs.
\end{enumerate}

The two objectives are not necessarily the same. The designers of Scratch note that for users ``who see programming as a medium for expression, not a path toward a career, Scratch is sufficient for their needs'' \cite{resnick2009scratch}.  The GP blocks environment is being designed to enable ``casual programmers'' to create ever more sophisticated programs, while removing limits that would force them to switch away from blocks \cite{GPBlocksAtYourFingertips2015}.

Before discussing the learning effects of blocks-based programming, we begin with a caution that it would be short-sighted to assume that tomorrow's programmers will program with the same languages and systems as today's. Each generation of programmers shifts the culture of coding, and the definition of ``real code'' will continue to evolve.

However, core learning questions remain. For students who 
continue with the study of traditional programming,
we can ask if a blocks-based introduction to programming is helpful or not.
This question has been directly tested in 
classrooms.


\subsection{Measuring Learning Transfer}

Research indicates that
learning a blocks language
can improve later learning of a traditional textual language. In a study of 10th graders learning C\# or Java \cite{Armoni:2015:SLP:2698235.2677087}, those who had taken a Scratch course in 9th grade learned
more quickly, understood loops better, and were more engaged and confident than their peers who had not.
However, in the final test,
a significant difference was seen in only one of three cognitive dimensions.
In a study
at two colleges \cite{moskal2004evaluating}, students with little or no previous programming experience and weak math preparation completed a CS0 programing class using 
Alice before beginning a Java CS1 course. 
Starting with Alice improved student grades (GPA of 3.0 versus 1.2 for non-Alice students) and the percentage of students taking further CS courses
(88\% compared to 47\%).

The question of whether such effects are simply due to giving students more preparation in an extra class has been tested by creating
courses that
combine a blocks-based introduction with a transition to a traditional language.
Reports from
courses using
Scratch before Java or C indicate
improved student engagement and understanding of some concepts \cite{MalanSIGCSE07, WolzStartingWithScratch}. In one study focused on learning transfer \cite{dann2012mediated},
an introductory Java course at CMU was modified to begin with Alice.
Students in this class that used both languages averaged 10\% or more better performance on every section of the same Java final exam, including expression evaluation, control structures, arrays, and working with class definitions.

That result is remarkable because one might assume that spending more time programming with blocks meant less time to learn Java.
The study used a
version of Alice that generated Java code from Alice
blocks, and
a \textit{mediated transfer}
pedagogy that
made explicit connections between programming concepts in Alice and Java.

Other studies of CS1 courses that switch from blocks to text without these features
have identified potential challenges to learning with blocks \cite{GarlickITICSE10,PowersSIGCSE07}.
Switching from a blocks language to text can involve both a change in syntax and semantics, and Shapiro and Ahrens propose teaching the transitions separately, teaching syntax before generalizing semantics \cite{shapiro2016}.
Additional research is needed to
identify the circumstances under which blocks are
effective.

Today, many introductory computer science courses use a blocks-before-text approach. In Harvard's CS50, students move from Scratch to C; Berkeley's CS10 progresses from Snap! to Python; 
Project Lead The Way's Computer Science Principles (CSP) course uses both Scratch and App Inventor before moving on to Python; and Code.org's CSP App Lab course moves from Droplet blocks to JavaScript.

\section{Why Blocks are Learnable}

In 2004, Ko, Myers and Aung \cite{ko2004six} identified six learning barriers encountered by non-programmers in programming tasks.
Three of these --- {\em selection}, {\em use}, and {\em coordination} --- reflect the difficulty of simply assembling a program. 

We believe the learnability of blocks languages arises from how they address the usability challenges underlying these three learning barriers:

\begin{enumerate}

\item Learning a programming vocabulary is hard. Blocks simplify this problem because picking a block from a palette is far easier than remembering a word: blocks rely on \textbf{recognition instead of recall}.

\item Code is hard to use because it presents a high cognitive load for new programmers. Blocks reduce the cognitive load by \textbf{chunking code} into a smaller number of meaningful elements.

\item Assembling code is error-prone. Blocks help users assemble code without basic errors by providing \textbf{constrained direct manipulation} of structure (e.g., two incompatible concepts do not have connecting parts).

\end{enumerate}



\begin{figure}
\centering
\subfigure[Scratch]{\includegraphics[height=0.85\columnwidth]{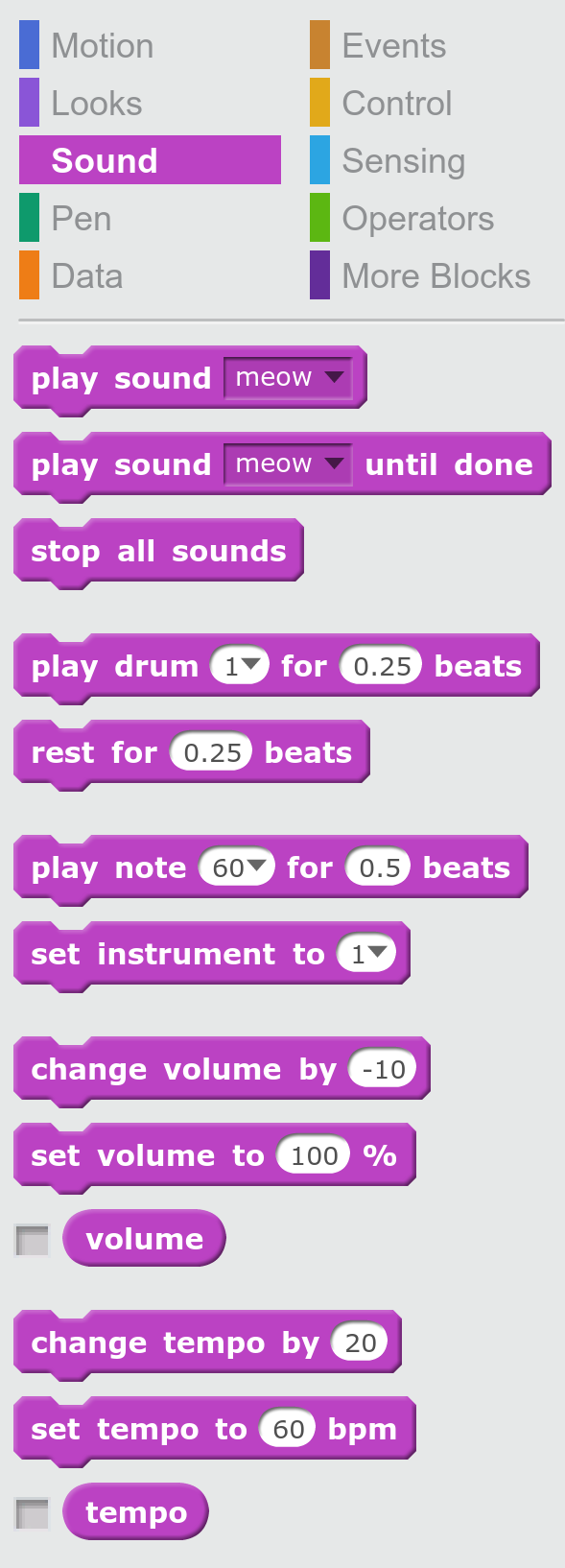}}
\hspace{0.03\columnwidth}
\subfigure[App Inventor]{\includegraphics[height=0.85\columnwidth]{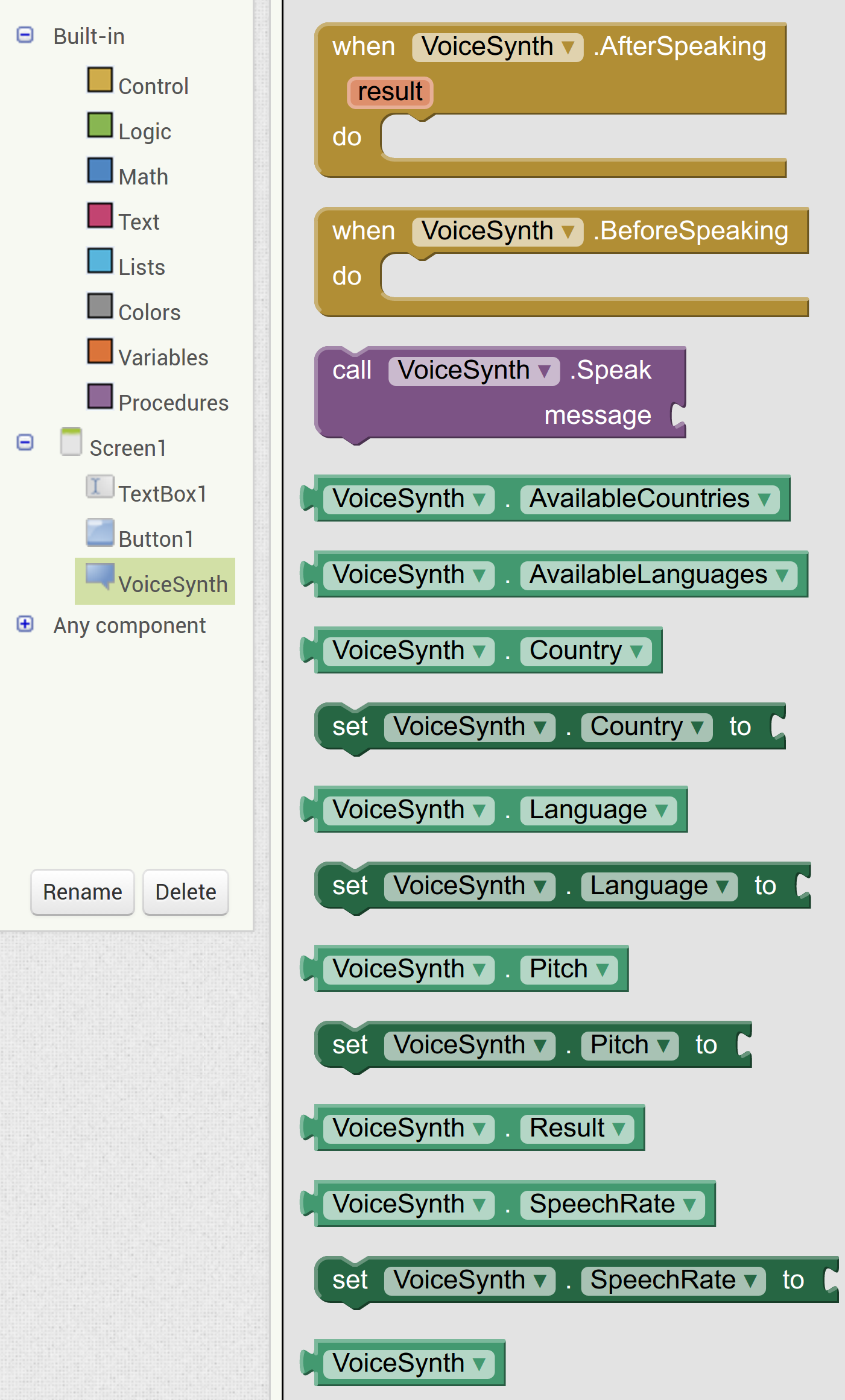}}
\vspace{-2mm}
\caption{The sound palette in Scratch and a voice synthesis palette in App Inventor. Palettes simplify the selection of programming elements by exploiting the ease of {\em recognition} over {\em recall}. Palettes organize concepts by topic, not name, and they remain open when used, allowing the user to discover and tinker with blocks based on their function.}
\label{palettes}
\end{figure}

\subsection{Recognition versus Recall}

Programming with a simple language or library typically involves a vocabulary of about 100-200 words. For example, HTML has 100 tags and 100 attributes, and SQL has about 200 keywords; Scratch is similar, with 130 blocks.  Recalling 100-200 concepts can overwhelm a newcomer.

Unlike text languages, blocks languages are intimately tied to their programming environments, and nearly all block environments have adopted a few interface conventions that address key usability problems. One such convention is tackling  vocabulary by organizing blocks in functionally related  palettes on the screen.

Palettes differ from autocomplete menus in professional code editors because they persist instead of disappearing and they organize concepts by topic instead of by name.  This design simplifies discovery and exploration.  Figure \ref{palettes}a shows the Sound palette in Scratch. It is an instructive reference showing all 13 methods for audio in that environment.

Similar organization is seen in larger blocks environments. To help manage the complexity of creating mobile apps, App Inventor provides a dynamic set of blocks, with additional blocks available in programs that have interactions with more components (Figure \ref{palettes}b). Yet, the basic idiom is the same as Scratch: explorable palettes organized by function.

Remembering the order, type, and valid values of operands is also daunting for newcomers. Many systems address this by supplying blocks with default operand values, drop-down menus and specialized editors to specify operands, and extra words to indicate operand meanings
(Figure \ref{playblock}a).

\subsection{Chunking Information with Blocks}
%
%
%

Programming languages present a high cognitive load to a student who is learning a new syntax. For example, consider the \texttt{for} loop in JavaScript syntax:

\begin{center}
\texttt{for (var i = 0; i < 50; i++) \{ $\cdots$ \} }
\end{center}

\noindent{}This dense notation is a barrier to beginners. 
In the words of one student, JavaScript ``is really confusing to understand with all the parentheses and brackets and all of that'' \cite{weintrop2015block}.

To understand the difficulty, consider that this code contains five words (\texttt{for var i i i}), ten pieces of punctuation (\texttt{( = ; < ; + + ) \{ \}} ), and two numbers (\texttt{0} and \texttt{50}), a total of seventeen units of information. Studies of human cognitive capacity have established that people have a working memory of about seven chunks of information \cite{miller1956magical}. Trying to understand this line of code as seventeen separate items may overwhelm the working memory of a new programmer.

%
%
%

\begin{figure}
\centering
\includegraphics[width=.9\columnwidth]{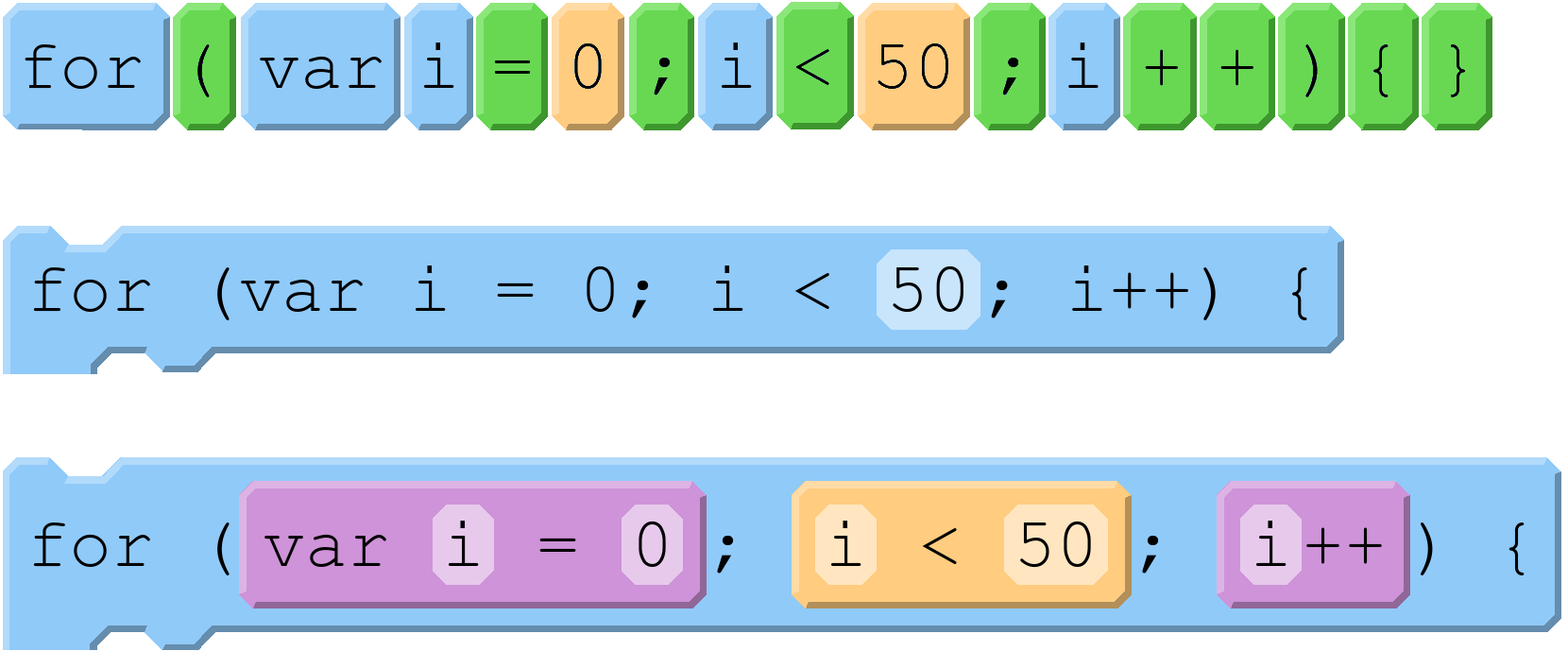}
\vspace{-2mm}
\caption{Three ways of reading a for loop in chunks. A naive reading of code (top) interprets the code as 17 chunks, but an expert reading of code (middle) interprets the most common form of loop as a single chunk, with a second chunk for the loop limit 50. An alternative (bottom) reading interprets three clauses as chunks. Code.org uses the middle rendering when introducing loops to high-school students for the first time, and switches to the bottom rendering when students are familiar with for loops.}
\label{threechunkings}
\end{figure}

Experienced JavaScript programmers have no problem understanding the line of code above because they have learned to interpret the code in larger chunks. Because a \texttt{for} loop follows a very common pattern, it can be read in just two chunks: first, the typical \texttt{for} loop that uses the conventional looping pattern (\texttt{i} starting at 0 and incrementing by 1); second, the particular choice of \texttt{50} as the upper limit. Figure \ref{threechunkings} illustrates different ways of chunking the code.

\begin{figure}
\centering
\subfigure[Scratch]{\includegraphics[width=0.5\columnwidth]{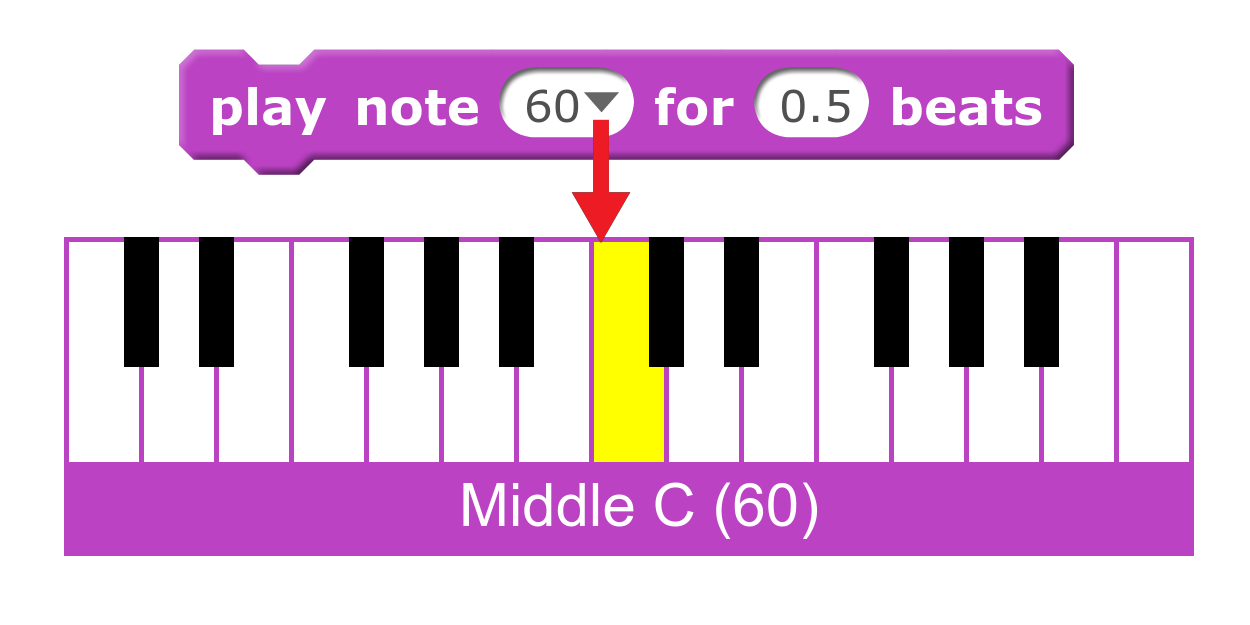}
}\quad
\subfigure[Python]{\raisebox{7mm}{\texttt{playNoteFor(60, 0.5)}
}}
\vspace{-3mm}
\caption{Blocks show structure visually instead of using punctuation. They can aid learnability using plain language, default values, and value pickers.}
\label{playblock}
\end{figure}

Blocks help reduce cognitive load by showing new programmers how to read larger chunks. In the Code.org Computer Science Principles course, blocks for JavaScript \texttt{for} loops are drawn just as an expert would see the code: as two chunks with a single block with a single socket for the loop upper bound. Complexity can also be reduced by nesting chunks within chunks. For example, Code.org reveals finer-grained \texttt{for} structure in more advanced portions of the same course, as illustrated at the bottom of Figure \ref{threechunkings}.

The use of blocks to chunk code aids readability even for simple commands, because blocks can forgo the punctuation that text code uses to denote structure and use explanatory words instead.  For example, as illustrated in Figure \ref{playblock}b, a simple call in Python requires reading delimiters and knowing argument order, whereas the equivalent block in Scratch reads naturally, using plain words to explain its behavior.

By organizing code as visible chunks, blocks help new programmers concentrate on what the code means rather than the notation that is used to write it.

\subsection{Direct Manipulation of Visible Structure}
The visual form of blocks alleviates the burden of assembling syntactically correct units by typing one character at a time. 
But there are other advantages of directly manipulating program fragments that have visual constraints.  

One benefit 
is that the blocks can help prevent errors by making the grammar of a program visible. Blocks can be seen as a form of syntax-directed editing with constrained direct manipulation. In 1981, creators of an early structured-editing tool noted, ``Programs are not text; they are hierarchical compositions of computational structures and should be edited, executed, and debugged in an environment that consistently acknowledges and reinforces this viewpoint'' \cite{teitelbaum1981}.

\begin{figure}
\centering
\includegraphics[trim={0 3mm 0 0},width=\columnwidth]{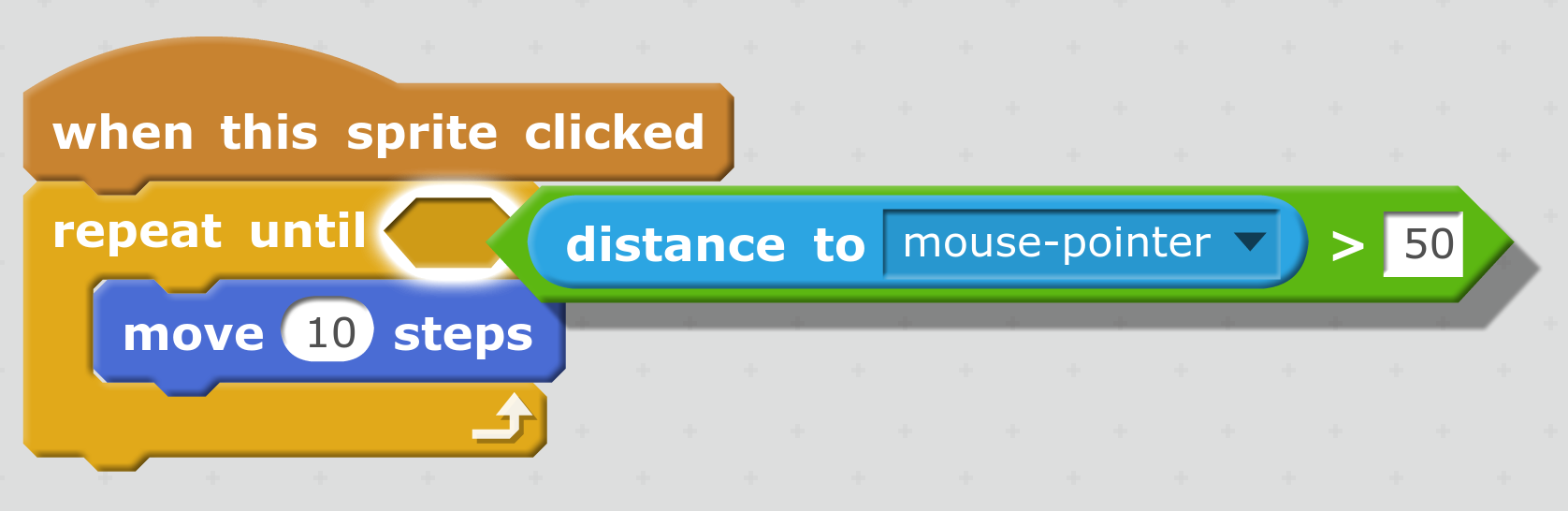}
\caption{Block shapes show and enforce rules of composition. Scratch commands compose vertically, and expressions fit into holes. Here, a Boolean expression (diamond shape) is being dropped into a matching 
hole for a loop test condition.}
\label{scratchshapes}
\end{figure}

Block shapes help beginners understand which grammatical phrases  (expresssions vs. commands vs. declarations) are legal in what contexts.  In Scratch, commands connect vertically with nubs and notches, whereas expressions are smooth shapes that fit into smooth holes. Constraints on drag-and-drop prevent the two types from being confused (Figure \ref{scratchshapes}). Students report that the puzzle shapes are helpful for assembling programs \cite{weintrop2015block}.

Visualization of types by shape can be applied to richer type systems: OpenBlocks provides 14 connector shapes to represent different types \cite{RoqueThesis}, and researchers have created experimental block languages with dynamically generated shapes to represent compositional type systems \cite{Lerner:2015:PBF:2702123.2702302,VasekHonorsThesis}. 

Directly manipulable blocks also encourage bottom-up tinkering with program pieces in ways not directly supported by raw text. Blocks programmers experiment with blocks by connecting them to build islands of code fragments on the programming surface that are isolated from the main program \cite{meerbaum2011habits,weintrop2015block}. In blocks environments supporting liveness \cite{maloney1995directness}, these fragments can be executed by pointing and clicking, providing a key benefit of interpreted text-based languages without a read-eval-print loop console separate from the editor. The program gradually grows as it is augmented by dropping in these fragments when they behave as desired. 

\subsection{Learnability Beyond Blocks}

Blocks aid in the construction of code, but blocks alone are not enough to make a programming language learnable. Users new to a language face additional learning challenges:

\begin{itemize}
\item They must wrestle with practical aspects like installing language tools, saving/loading programs, etc.;
\item They must learn the vocabulary of the language and understand the concepts denoted by its words;
\item They need to understand runtime semantics such as  flow of control and changes in state over time;
\item They eventually need to learn common patterns of use, moving beyond isolated concepts.
\end{itemize}

Each of these learning hurdles can be helped or hindered by the programming environment, and each of these problems is an area of active research and development.

\subsubsection{Programming Online}

To simplify installation, programming tools are moving online. When a programming environment is in a web browser,
a new programmer is just a few clicks away from creating a first program.
A cloud-based programming tool can provide a complete and consistent programming environment with fewer potential problems. 

Although blocks programming environments for beginners have long been offered online, text-based programming environments are also becoming available online. With tools such as Cloud 9, CodeAnywhere, and CodeEnvy, programmers of all levels can benefit from working online.

\subsubsection{Words, Concepts, and Abstractions}

The names chosen for language constructs can have an impact on learnability. Empirical studies by Stefik and Seibert 
\cite{Stefik:2013:EIP:2543488.2534973} have found that common keywords like \texttt{for} or operators such as \texttt{!=} serve as hard-to-learn jargon, and are not as easily learnable as familiar words such as \texttt{repeat} or \texttt{unequal}. They found that the syntax of 
languages such as Java and Perl are no more learnable than a synthetic programming language with randomly-selected punctuation used for keywords.

Even more important for learnability are the abstractions chosen by language designers to allow users to build simple programs with compelling behavior.
For example, the designers of Alice
worked with users to
develop intuitive abstractions for controlling 3D animations. During their design process, the Alice team eliminated
jargon such as transformation matrices and substituted more intuitive concepts like object-relative motions. These new abstractions made it
easier for users to specify 3D animations \cite{Conway-SIGCSE2000}.


%

Designing languages and libraries focused on learnability based on empirical evidence is a major area for future work.

\subsubsection{Runtime Understanding}

The dynamic state of a program can be made more understandable by making its state visible. For example, Code.org highlights the individual block that is actively running so the correspondence between code and action can be seen.  Snap! provides widgets for every variable to show the current state.

Even with highly visible state, understanding actions in the past or future can be difficult. Liveness \cite{maloney1995directness} is one approach to addressing this problem. A live system aims to make actions concrete by applying them immediately to the current state. 
For example, in App Inventor, Scratch, and Snap!, many edits to a running block program take effect immediately, without the need to restart the program. 


Another approach for making the evolution of state understandable is to allow the programmer to travel in time by inspecting, advancing or rewinding the timeline of a program. The concept of omniscient debugging was first described in the early 1970s as a capability to trace backward in time through  execution history to identify the location of the fault that caused a later observed failure \cite{zelkowitz1973}. TRAKLA2 \cite{nikander2004visual}, UUhistle \cite{sorva2011context} and Online Python Tutor \cite{guo2013online} provide this capability for Java and Python.


\subsubsection{Examples and Reuse}


The activity of programming has changed with the availability of large repositories containing examples of shared code. Programmers of all levels report finding and adapting examples as a core programming activity
\cite{Brandt:2009:TSO:1518701.1518944,Dorn:2006:GDP:1151588.1151608}. In response, professional programming and end-user programming environments are beginning to incorporate example support \cite{Brandt:2010:EPI:1753326.1753402, Sawadsky:2011:FTC:1984708.1984722}.
Novices, too, want to learn via examples, but may struggle to do so
\cite{Rosson_Ballin_Nash_Everyday_Programming_2004}


Blocks-based languages like Scratch and Looking Glass use online sharing and remixing programs to provide example access.
But there is a tradeoff between the simplicity of reuse and the robustness of reused code. For example, Scratch simplifies sharing of code examples for novices by providing a ``backpack'' for collecting program snippets and assets that can be shared and dragged into a new project.
However, the backpack does not necessarily guarantee that the code will run correctly in a new project.

In contrast, Looking Glass uses a more complex process for reuse in which users select the beginning and end of behavior they want to use. Coupled with execution history information, this can ensure that the selected code will function within the context of a new program. A Play \& Explore feature allows users to connect program output to the line or lines of code that caused it, helping users to understand and begin to modify reused code.


%
%
%
\section{Scaling Blocks Code}
Why don't professionals program with block interfaces? One reason is that direct manipulation has efficiency disadvantages when making small edits. When creating an expression such as $(a / 2 + b / 2)$ in blocks, the programmer must find and drag blocks for each of the three arithmetic operators, and then fill in holes with variables and numbers.  Similarly, when rearranging an expression from $(a / 2 + b / 2)$ to $(a + b) / 2$, the expression tree needs to be pulled apart and put together again, requiring more gestures and more forethought than making the edits in text. HCI researchers observed that visual programming languages can have a higher \textit{viscosity} than text code because they make small changes harder \cite{Green1989}.

\begin{figure}
\centering
\includegraphics[trim={0 3mm 0 0},width=0.98\columnwidth]{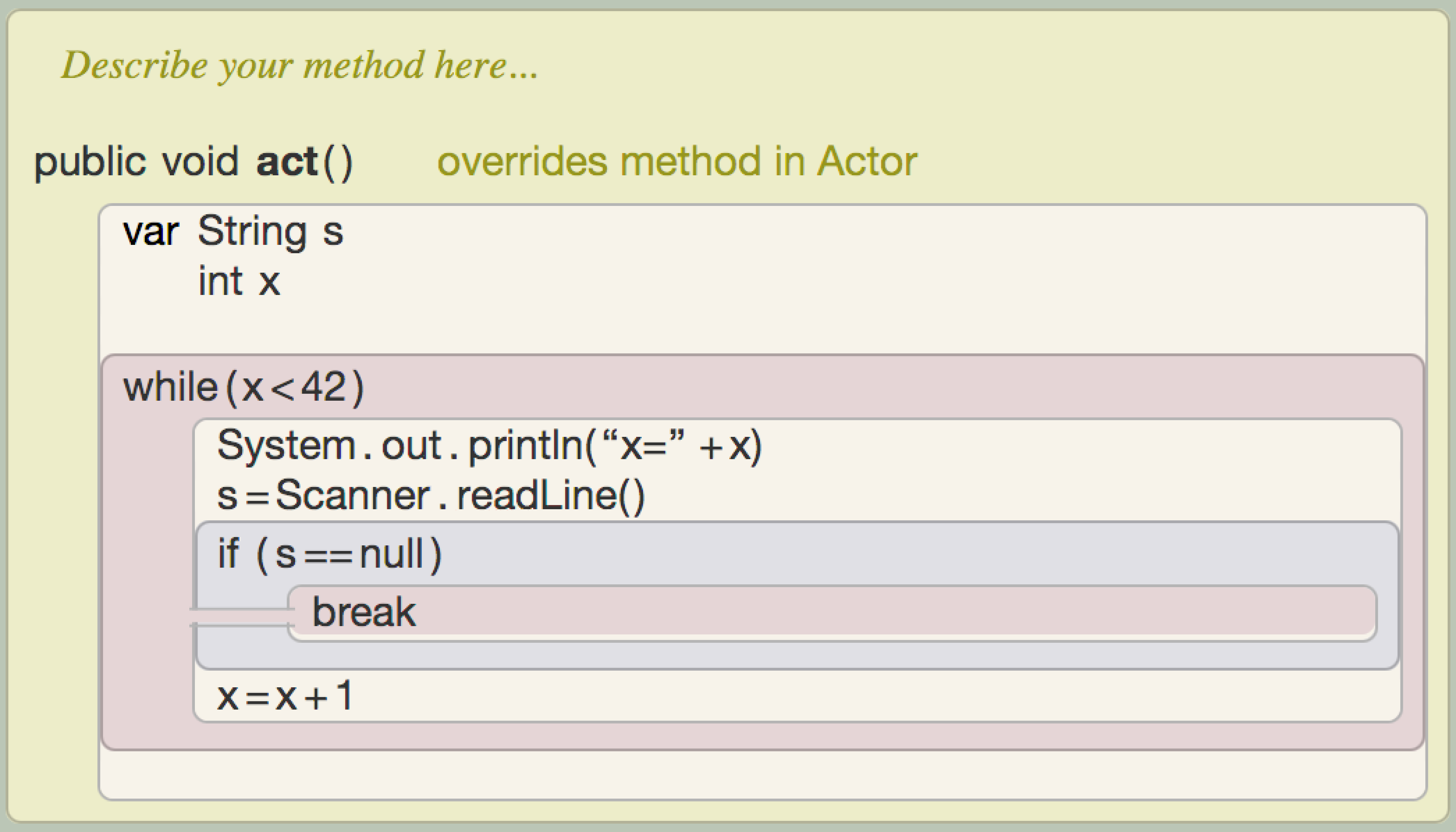}
\caption{Greenfoot's Stride editor combines text-style editing for expression-level details with drag-and-drop blocks for higher-level program structure.}
\label{greenfoot}
\end{figure}

Beyond viscosity, blocks environments can have several other usability disadvantages compared to textual programming languages:

\begin{itemize}
\item Low density: blocks take more space on the screen than equivalent text code.
\item Search and navigation: it can be challenging to find and navigate to the relevant part of a blocks program in a 2D workspace, only part of which may be visible. 
\item Source control: collaboration and version control systems are difficult to use without a text representation.
\end{itemize}

The newest generation of blocks programming tools include features that are designed to resolve the tension between usability advantages of text versus blocks.  There are two approaches: text-style entry and bidirectional mode switching.

\subsection{Text-Style Entry of Blocks}

Some new blocks environments, such as Greenfoot's frame-based Stride editor \cite{GreenfootFrameBasedEditing} and GP \cite{GPBlocksAtYourFingertips2015}, are designed to be used by programmers to create large programs, so efficient editing is an important design goal.

Both Stride and GP improve efficiency by providing text-based editing shortcuts within a blocks-oriented interface. To allow users to circumvent the step of finding a block on the palette, these systems let programmers enter blocks through an in-line autocomplete mechanism. Blocks can still be chosen from a palette, but a knowledgeable programmer can insert them by typing. 

The Stride editor also introduces a hybrid approach to editing code, differentiating between low-level and high-level structure (Figure \ref{greenfoot}). For expression-level code, it hides syntactic structure and allows traditional text editing, providing high-density display and lower viscosity. Visible tree structure and drag-and-drop manipulation are used for higher-level code such as control flow and class declarations.

\subsection{Bidirectional Mode Switching}

\begin{figure}
\centering
\includegraphics[trim={0 5mm 0 0}, width=0.95\columnwidth]{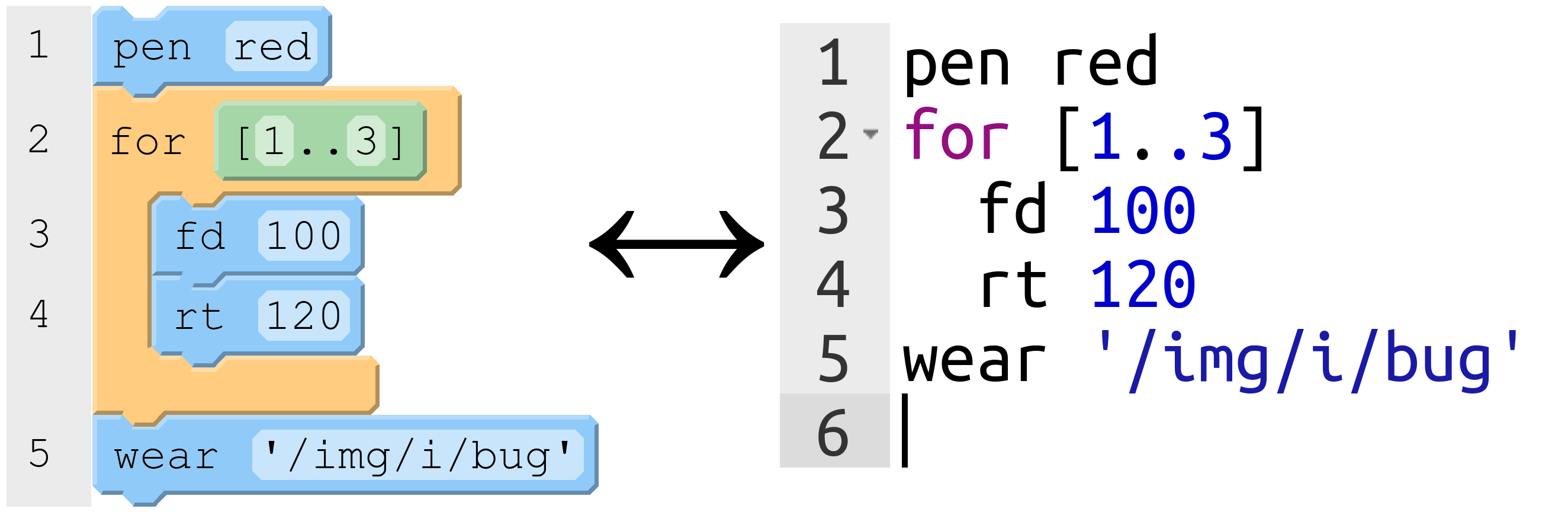}
\caption{Pencil Code provides bidirectional switching between blocks and text. Mode switching allows users to learn with blocks and edit quickly with text.}
\label{transition}
\end{figure}

Some blocks environments provide a bidirectional transformation between a traditional text language and a blocks representation of that language. These include 
Pencil Code (Coffee Script) \cite{PencilCodeIDC} (Figure \ref{transition}), Code.org's App Lab\footnote{\url{https://code.org/educate/applab}} (JavaScript), BlockEditor (Java) \cite{Matsuzawa-SIGCSE2015}, and Tiled Grace (Grace) \cite{Homer_Noble_Tiled_Grace_VISSOFT_2014}. Alice and Blockly provide non-editable views of text code.

The hypothesis that motivates the design of dual-mode tools is that users may benefit from the learnability of blocks in one mode, while they learn syntax and get the efficiency of text in the other mode.  This goal requires that the views be linked. For text to be safe for users who may want to return to blocks, it must be possible to return to blocks.

In dual-mode editors, the text code is the primary representation of the program, and blocks are a projected user interface view derived by parsing. This approach allows the editor to fully represent text information such as spacing, but it also means that the editor must allow syntax errors that would not have been possible in blocks. Error-recovery heuristics convert simple text syntax errors to special error blocks, but complex errors can prevent mode switching.

\subsection{Comparing Approaches}

There is a tradeoff between the two approaches to unifying blocks and text. While the dual-mode editors provide direct support for learning traditional text syntax such as JavaScript or Java, they also impose the cognitive overhead of working with syntax errors that that can only be introduced in text mode. Visualization research on multiple coordinated views suggests that the benefits of providing more than one view need to be balanced against the cognitive overhead imposed by switching between views \cite{wang2000guidelines}.

Single mode structured editors have the advantage of a conceptual model free of syntax errors, because the primary object being edited is the abstract syntax tree. However, to maintain consistency, many kinds of textual edits must be prevented, and other edits may require special tree editing commands. These constraints raise editing viscosity, and they may present additional cognitive barriers.

Interfaces that effectively bridge the gap between blocks and text are an active area of research.


%
%
%
%
%
\section{Applying Blocks: Two Examples}

Let us take a look at experimental blocks languages in two specific domains that are unfamiliar to most programmers.

%
%
%

\subsection{Programming 3D Printers}

\begin{figure}
\centering
\fbox{\includegraphics[width=0.95\columnwidth]{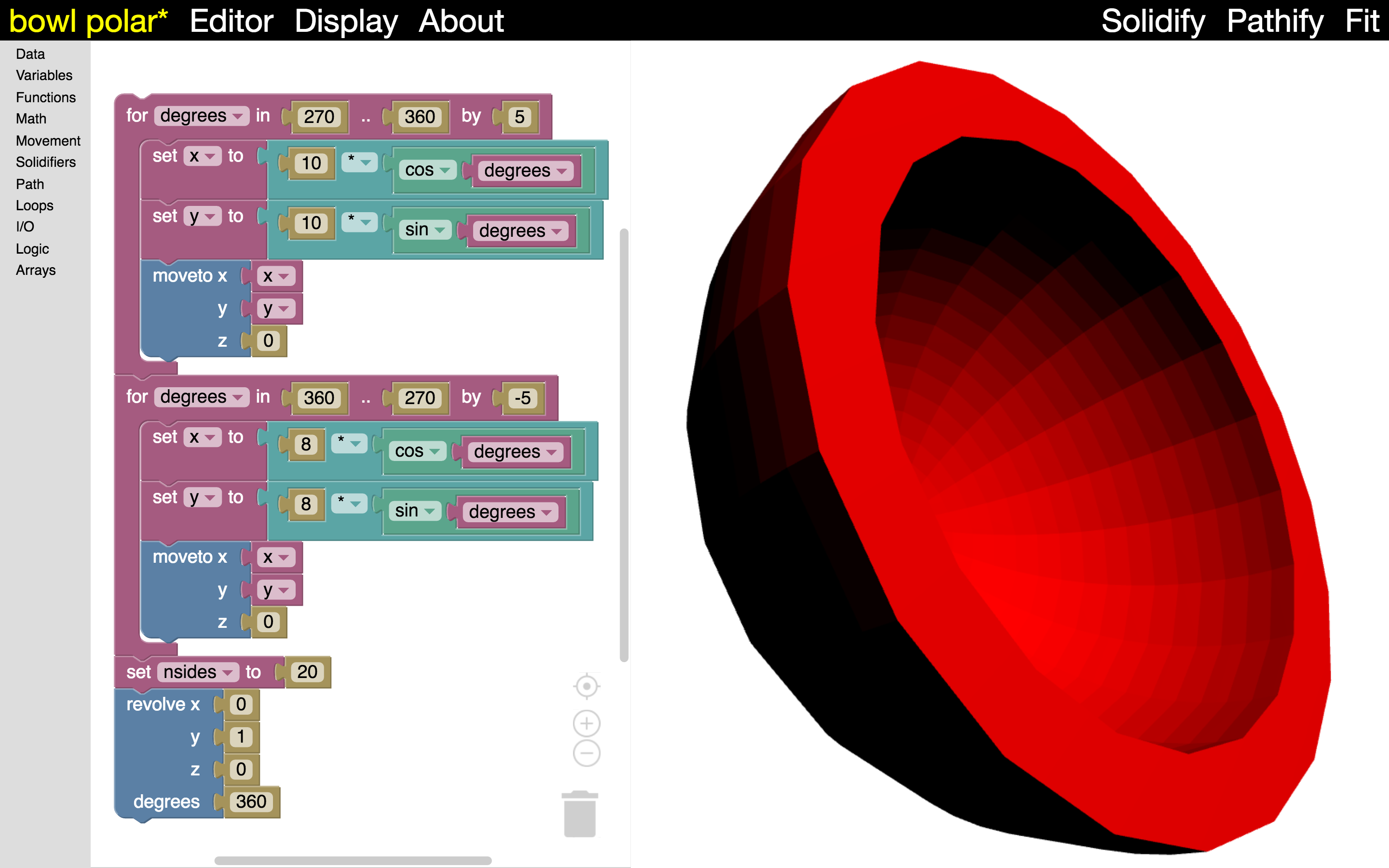}}
\caption{Blocks programming in MadeUp. 3D printing is an area of rapid innovation, and blocks make it possible to use new 3D modeling languages without a steep learning curve.}
\label{madeupbowl}
\end{figure}

Traditionally, 3D printing models are drawn interactively 
by direct manipulation using CAD software. 
Writing custom code to create a model is a powerful alternative approach, 
but coding for 3D fabrication has traditionally been 
the province of a few expert programmers. 
Now, the falling cost and rising availability of 3D printers 
has made it possible for non-specialists to write their own custom 3D modeling code.

Two recently developed languages bring programming for 3D printing 
to novice users: BeetleBlocks \cite{BeetleBlocksIGCC2012} 
and Madeup \cite{MadeupBB}.
Although the two systems have different languages, 
there are several commonalities: both are web-based interfaces 
with a live 3D rendering of the shape being created, 
and both provide a blocks language to simplify learning.
BeetleBlocks follows the principles of turtle graphics: a beetle moves a 
``pen" that can be turned on and off,
and 3D shapes can be created out of iterated strokes.
MadeUp (Figure \ref{madeupbowl}) takes a more abstract approach, allowing users to 
trace out both paths and parametric surfaces. 
Special functions can
rotate or extrude the paths to create solids.

The two languages offer different levels of power and abstraction. 
Which language is the right one to use? This domain is an excellent 
example of the advantage of the learnability of blocks. 
Both blocks languages have a very shallow learning curve, 
and it is easy to try both.

\subsection{Querying the Semantic Web}

Another domain where learnability is essential is in querying large datasets. 
Consider the problem of querying Resource Description Framework
(RDF) data from the semantic web. 
SPARQL is the standard language for querying RDF; 
it includes several constructs for working with RDF triples 
that distinguish it from other query languages such as SQL. 
However, potential users of SPARQL face two hurdles: First, programmers must learn the vocabulary and syntax of SPARQL, with its specialized operators and constructs. Second, querying RDF requires not just knowledge of the language, but also knowledge of instance and schema data.

To address both problems, Paolo Bottoni and Miguel Ceriani 
have created a blocks language they call 
the \textit{SPARQL Playground} \cite{BottoniCerianiBB}.
Rather than helping users learn about sequential programs, their blocks language helps users select, filter, and join data using SPARQL primitives.

\begin{figure}
\centering
\includegraphics[trim={0 3mm 0 0},width=.85\columnwidth]{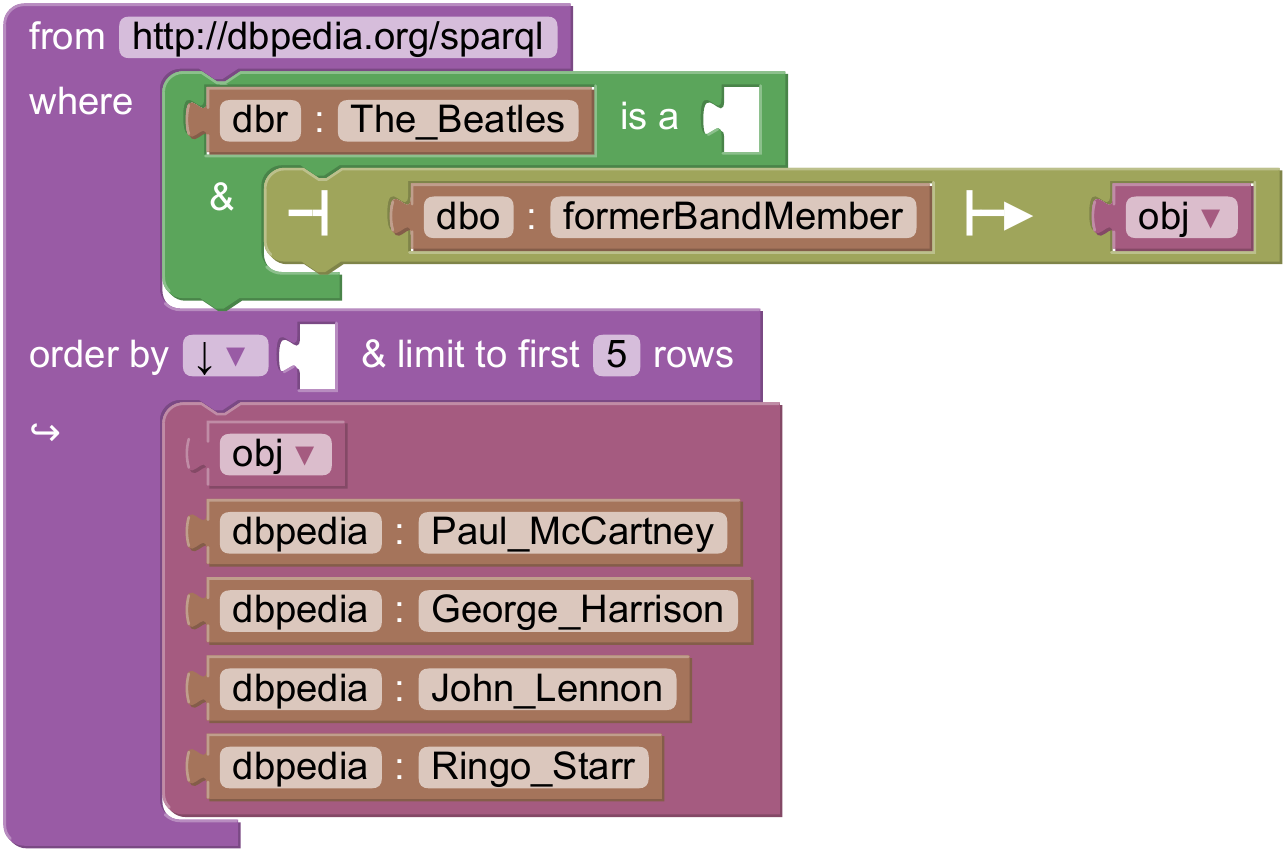}
\caption{The SPARQL Playground is a blocks-based query execution tool that provides blocks for constructing queries of RDF data. Query results (bottom) are also provided as blocks, and they can be dragged to build into other queries.}
\label{sparqlquery}
\end{figure}

The SPARQL playground is interesting for a second reason: all query results in the playground are also returned as draggable blocks (Figure \ref{sparqlquery}).
This feature allows users to save instance data on the programming workspace 
to be incorporated into new queries. With the SPARQL playground,
it is easy to begin with general queries to explore the types of data available, 
then use these discoveries to make refinements.

\section{Making New Blocks Languages}

%
%
%
It is now possible to create your own domain-specific blocks
environment using a blocks-based language toolkit. 
Blocks-language authors should be aware of at least three toolkits: Blockly\footnote{\url{https://developers.google.com/blockly}},
Droplet \cite{DropletCCSCNE}, 
and OpenBlocks \cite{RoqueThesis}.

The original blocks metalanguage is MIT OpenBlocks. Created in 2007 by Ricarose Roque as the basis of StarLogo TNG, it allows a large degree of geometric customization.  
OpenBlocks has also been used in App Inventor Classic  
and BlockEditor. One drawback of OpenBlocks is that it requires users to download and install the Java JDK.

The challenge of installation is addressed by Blockly, an HTML-based block language toolkit by Neil Fraser of Google. 
Blockly is currently the most popular blocks language toolkit: 
it is the tool behind App Inventor, the SPARQL playground, and MadeUp, as well as the Code.org Hour of Code puzzles. 
Future versions of Scratch will also use Blockly. 

Droplet is the newest of the blocks language creation toolkits, 
developed by Anthony Bau for use in Pencil Code, and also 
used by Code.org in their App Lab. 
It is newer and less mature than OpenBlocks and Blockly, 
but 
takes a unique approach that allows seamless bidirectional transformation 
between all blocks and textual code. 

\section{Summary}

When a programming language is provided as a user interface that welcomes novice users, rather than as a technical tool only for experienced developers, we arrive at a new picture of what the programming environment should do:
\begin{itemize}
\setlength{\itemsep}{0pt}
\item Vocabulary should derive from recognition, not recall;
\item Cognitive load should be lowered by chunking code;
\item Grammar rules and types should be made visible;
\item Program chunks should be directly manipulable;
\item Low-viscosity editing should also be possible;
\item Coding should work without installation of tools;
\item Simple concepts should be described with clear words and high-level abstractions;
\item Runtime state and behavior should be visible;
\item Examples should be easy to find and apply.
\end{itemize}
\noindent
In short, for a programming tool to be usable by new or casual programmers, its design must focus on learnability.

Blocks have proven to be effective at solving many of these problems. Although programming is still not nearly as widely learned as it should be, the progress made by blocks language interfaces can inspire us all to see that programming can be made more learnable.

The art of programming is the original human-computer interaction, and it remains an unsolved usability challenge. We can still do more to make programming available to all.


\bibliographystyle{abbrv}
\bibliography{blocks-and-beyond}  

\begin{thebibliography}{10}

\bibitem{Armoni:2015:SLP:2698235.2677087}
M.~Armoni, O.~Meerbaum-Salant, and M.~Ben-Ari.
\newblock From {S}cratch to ``real" programming.
\newblock {\em Transactions on Computing Education}, 14(4), Feb. 2015.

\bibitem{PencilCodeIDC}
D.~Bau, D.~A. Bau, M.~Dawson, and C.~S. Pickens.
\newblock Pencil {C}ode: Block code for a text world.
\newblock In {\em 14th International Conference on Interaction Design and
  Children}, pages 445--448, 2015.

\bibitem{DropletCCSCNE}
D.~A. Bau.
\newblock Droplet, a blocks-based editor for text code.
\newblock {\em Journal of Computing Sciences in Colleges}, 30(6):138--144, June
  2015.

\bibitem{BottoniCerianiBB}
P.~Bottoni and M.~Ceriani.
\newblock Using blocks to get more blocks: Exploring linked data through
  integration of queries and result sets in block programming.
\newblock In {\em IEEE Blocks and Beyond Workshop}, pages 99--102, Oct. 2015.

\bibitem{Brandt:2010:EPI:1753326.1753402}
J.~Brandt, M.~Dontcheva, M.~Weskamp, and S.~R. Klemmer.
\newblock Example-centric programming: Integrating web search into the
  development environment.
\newblock In {\em ACM SIGCHI Conference on Human Factors in Computing Systems},
  CHI '10, pages 513--522, 2010.

\bibitem{Brandt:2009:TSO:1518701.1518944}
J.~Brandt, P.~J. Guo, J.~Lewenstein, M.~Dontcheva, and S.~R. Klemmer.
\newblock Two studies of opportunistic programming: Interleaving web foraging,
  learning, and writing code.
\newblock In {\em ACM SIGCHI Conference on Human Factors in Computing Systems},
  CHI '09, pages 1589--1598, 2009.

\bibitem{Conway-SIGCSE2000}
M.~Conway, S.~Audia, T.~Burnette, D.~Cosgrove, and K.~Christiansen.
\newblock Alice: Lessons learned from building a {3D} system for novices.
\newblock In {\em ACM SIGCHI Conference on Human Factors in Computing Systems},
  CHI '00, pages 486--493, 2000.

\bibitem{dann2012mediated}
W.~Dann, D.~Cosgrove, D.~Slater, D.~Culyba, and S.~Cooper.
\newblock Mediated transfer: {A}lice 3 to {J}ava.
\newblock In {\em 43rd ACM Technical Symposium on Computer Science Education},
  SIGCSE '12, pages 141--146, 2012.

\bibitem{Dorn:2006:GDP:1151588.1151608}
B.~Dorn and M.~Guzdial.
\newblock Graphic designers who program as informal computer science learners.
\newblock In {\em Second International Workshop on Computing Education
  Research}, ICER '06, pages 127--134, 2006.

\bibitem{GarlickITICSE10}
R.~Garlick and E.~C. Cankaya.
\newblock Using {A}lice in {CS1}: a quantitative experiment.
\newblock In {\em 15th Annual Conference on Innovation and Technology in
  Computer Science Education}, ITiCSE '10, pages 165--168, 2010.

\bibitem{Green1989}
T.~R.~G. Green.
\newblock Cognitive dimensions of notations.
\newblock In A.~Sutcliffe and L.~Macaulay, editors, {\em People and Computers
  V}, pages 443--460. Cambridge University Press, Cambridge, UK, 1989.

\bibitem{guo2013online}
P.~J. Guo.
\newblock Online {P}ython tutor: Embeddable web-based program visualization for
  {CS} education.
\newblock In {\em 44th ACM Technical Symposium on Computer Science Education},
  SIGCSE '13, pages 579--584, 2013.

\bibitem{Homer_Noble_Tiled_Grace_VISSOFT_2014}
M.~Homer and J.~Noble.
\newblock Combining tiled and textual views of code.
\newblock In {\em 2014 Second IEEE Working Conference on Software
  Visualization}, VISSOFT '14, pages 1--10, Sept 2014.

\bibitem{MadeupBB}
C.~Johnson and P.~Bui.
\newblock Blocks in, blocks out: A language for {3D} models.
\newblock In {\em IEEE Blocks and Beyond Workshop}, pages 77--82, Oct. 2015.

\bibitem{ko2004six}
A.~J. Ko, B.~A. Myers, and H.~H. Aung.
\newblock Six learning barriers in end-user programming systems.
\newblock In {\em IEEE Symposium on Visual Languages and Human Centric
  Computing}, VL/HCC '04, pages 199--206, 2004.

\bibitem{GreenfootFrameBasedEditing}
M.~K\"{o}lling, N.~C.~C. Brown, and A.~Altadmri.
\newblock Frame-based editing: Easing the transition from blocks to text-based
  programming.
\newblock In {\em 10th Workshop in Primary and Secondary Computing Education},
  WiPSCE '15, Nov. 2015.

\bibitem{BeetleBlocksIGCC2012}
D.~Koschitz and E.~Rosenbaum.
\newblock Exploring algorithmic geometry with ``{B}eetle {B}locks:'' a
  graphical programming language for generating 3d forms.
\newblock In {\em 15th International Conference on Geometry and Graphics},
  pages 380--389, Aug. 2012.

\bibitem{Lerner:2015:PBF:2702123.2702302}
S.~Lerner, S.~R. Foster, and W.~G. Griswold.
\newblock Polymorphic blocks: Formalism-inspired {UI} for structured
  connectors.
\newblock In {\em 33rd Annual ACM Conference on Human Factors in Computing
  Systems}, CHI '15, pages 3063--3072, 2015.

\bibitem{MalanSIGCSE07}
D.~J. Malan and H.~H. Leitner.
\newblock Scratch for budding computer scientists.
\newblock {\em ACM SIGCSE Bulletin}, 39(1):223--227, Mar. 2007.

\bibitem{maloney1995directness}
J.~H. Maloney and R.~B. Smith.
\newblock Directness and liveness in the morphic user interface construction
  environment.
\newblock In {\em 8th Annual ACM Symposium on User Interface and Software
  Technology}, pages 21--28, 1995.

\bibitem{Matsuzawa-SIGCSE2015}
Y.~Matsuzawa, T.~Ohata, M.~Sugiura, and S.~Sakai.
\newblock Language migration in non-{CS} introductory programming through
  mutual language translation environment.
\newblock In {\em 46th ACM Technical Symposium on Computer Science Education},
  SIGCSE '15, pages 185--190, 2015.

\bibitem{meerbaum2011habits}
O.~Meerbaum-Salant, M.~Armoni, and M.~Ben-Ari.
\newblock Habits of programming in {S}cratch.
\newblock In {\em 16th Annual Joint Conference on Innovation and Technology in
  Computer Science Education}, ITiCSE '11, pages 168--172, 2011.

\bibitem{miller1956magical}
G.~A. Miller.
\newblock The magical number seven, plus or minus two: some limits on our
  capacity for processing information.
\newblock {\em Psychological Review}, 63(2):81, 1956.

\bibitem{GPBlocksAtYourFingertips2015}
J.~M\"{o}nig, Y.~Ohshima, and J.~Maloney.
\newblock Blocks at your fingertips: Blurring the line between blocks and text
  in {GP}.
\newblock In {\em IEEE Blocks and Beyond Workshop}, pages 51--53, Oct. 2015.

\bibitem{moskal2004evaluating}
B.~Moskal, D.~Lurie, and S.~Cooper.
\newblock Evaluating the effectiveness of a new instructional approach.
\newblock {\em ACM SIGCSE Bulletin}, 36(1):75--79, 2004.

\bibitem{nikander2004visual}
J.~Nikander, A.~Korhonen, O.~Sepp{\"a}l{\"a}, V.~Karavirta, P.~Silvasti, and
  L.~Malmi.
\newblock Visual algorithm simulation exercise system with automatic
  assessment: Trakla2.
\newblock {\em Informatics in Education-An International Journal},
  3(2):267--288, 2004.

\bibitem{PowersSIGCSE07}
K.~Powers, S.~Ecott, and L.~M. Hirshfield.
\newblock Through the looking glass: Teaching {CS0} with {A}lice.
\newblock {\em ACM SIGCSE Bulletin}, 39(1):213--217, Mar. 2007.

\bibitem{resnick2009scratch}
M.~Resnick, J.~Maloney, A.~Monroy-Hern{\'a}ndez, N.~Rusk, E.~Eastmond,
  K.~Brennan, A.~Millner, E.~Rosenbaum, J.~Silver, B.~Silverman, and Y.~B.
  Kafai.
\newblock Scratch: programming for all.
\newblock {\em Communications of the ACM}, 52(11):60--67, 2009.

\bibitem{RoqueThesis}
R.~Roque.
\newblock Openblocks: An extendable framework for graphical block programming
  systems.
\newblock Master's thesis, MIT, May 2007.

\bibitem{Rosson_Ballin_Nash_Everyday_Programming_2004}
M.~B. Rosson, J.~Ballin, and H.~Nash.
\newblock Everyday programming: Challenges and opportunities for informal web
  development.
\newblock In {\em 2004 IEEE Symposium on Visual Languages and Human Centric
  Computing}, VL/HCC'04, pages 123--130, Sept 2004.

\bibitem{Sawadsky:2011:FTC:1984708.1984722}
N.~Sawadsky and G.~C. Murphy.
\newblock Fishtail: From task context to source code examples.
\newblock In {\em 1st Workshop on Developing Tools As Plug-ins}, pages 48--51,
  2011.

\bibitem{shapiro2016}
R.~B. Shapiro and M.~Ahrens.
\newblock Beyond blocks: Syntax and semantics.
\newblock {\em Communications of the ACM}, 59(5):39--41, 2016.

\bibitem{sorva2011context}
J.~Sorva and T.~Sirki{\"a}.
\newblock Context-sensitive guidance in the {UU}histle program visualization
  system.
\newblock In {\em Proceedings of the 6th Program Visualization Workshop}, pages
  77--85, 2011.

\bibitem{Stefik:2013:EIP:2543488.2534973}
A.~Stefik and S.~Siebert.
\newblock An empirical investigation into programming language syntax.
\newblock {\em Transactions on Computing Education}, 13(4):19:1--19:40, Nov.
  2013.

\bibitem{teitelbaum1981}
T.~Teitelbaum and T.~Reps.
\newblock The {C}ornell program synthesizer: A syntax-directed programming
  environment.
\newblock {\em Communications of the ACM}, 24(9):563--573, 1981.

\bibitem{VasekHonorsThesis}
M.~Vasek.
\newblock Representing expressive types in blocks programming languages.
\newblock Undergraduate thesis, Wellesley College, May, 2012.

\bibitem{wang2000guidelines}
M.~Q. Wang~Baldonado, A.~Woodruff, and A.~Kuchinsky.
\newblock Guidelines for using multiple views in information visualization.
\newblock In {\em Proceedings of the Working Conference on Advanced Visual
  Interfaces}, pages 110--119, 2000.

\bibitem{weintrop2015block}
D.~Weintrop and U.~Wilensky.
\newblock To block or not to block, that is the question: students' perceptions
  of blocks-based programming.
\newblock In {\em 14th International Conference on Interaction Design and
  Children}, pages 199--208, 2015.

\bibitem{WolzStartingWithScratch}
U.~Wolz, H.~H. Leitner, D.~J. Malan, and J.~Maloney.
\newblock Starting with {S}cratch in {CS1}.
\newblock {\em ACM SIGCSE Bulletin}, 41(1):2--3, Mar. 2009.

\bibitem{zelkowitz1973}
M.~Zelkowitz.
\newblock Reversible execution.
\newblock {\em Communications of the ACM}, 16(9):566, 1973.

\end{thebibliography}
%


\section*{APPENDIX: SIDEBAR}

\section*{Blocks-Based Educational Tools}

\noindent\textit{On the web}:
\begin{itemize}
\item Scratch (https://scratch.mit.edu) - in-browser animation and game creation, with support for extensions
\item Code.org (http://code.org) - a variety of tools including puzzle programming exercises with tutorial videos
\item Snap!~(http://snap.berkeley.edu) - enhanced language inspired by Scratch that includes first class functions
\item App Inventor (http://appinventor.mit.edu) - creation of Android apps using in-browser blocks IDE
\item Pencil Code (https://pencilcode.net) - creates CoffeeScript web apps, transforming between text and blocks
\item StarLogo Nova (http://www.slnova.org) - multi-agent simulations and games in a 3D rendered world
\item Blockly Games (https://blockly-games.appspot.com/) - a set of puzzles to solve with programming blocks
\item GameBlox (https://gameblox.org) - game creation that includes clonable agents, physics, and more
\end{itemize}

\noindent\textit{Downloadable:}
\begin{itemize}
\item AgentSheets / AgentCubes (http://www.agentsheets.com) - pioneering blocks environments for creating rule-based games and simulations

\item Alice (http://www.alice.org) - pioneering blocks environment for creating 3D virtual worlds; support for exporting to Java
%
%
%
\item Looking Glass (https://lookingglass.wustl.edu) - 3D animated story creation; supports independent learning

\item  Kodu (http://www.kodugamelab.com) - rule-based programming of games for xBox and PC
\end{itemize}

\noindent\textit{On mobile:}
\begin{itemize}
\item Scratch Jr (http://www.scratchjr.org) - programming of animated scenes aimed at preliterate children
\item Pocket Code (http://www.catrobat.org) - blocks programming for small form factor of mobile devices
\item Tynker (https://www.tynker.com) - polished commercial platform for game and animation creation
\item Hopscotch (https://www.gethopscotch.com) - creating games and animations on iPhone and iPad
\end{itemize}

\end{document}